\documentclass{article}

\usepackage[preprint]{neurips_2024}
\usepackage{colortbl}
\usepackage[table]{xcolor}

\usepackage{hyperref}
\hypersetup{
    colorlinks=true,
    linkcolor=blue,
    filecolor=magenta,      
    urlcolor=cyan,
    citecolor=green,
    pdftitle={YingMusci-Singer},
    pdfauthor={GiantAILab}
}

\usepackage[utf8]{inputenc} % allow utf-8 input
\usepackage[T1]{fontenc}    % use 8-bit T1 fonts
\usepackage{hyperref}       % hyperlinks
\usepackage{url}            % simple URL typesetting
\usepackage{booktabs}       % professional-quality tables
\usepackage{multirow}        % for multirow cells in tables
\usepackage{amsfonts}       % blackboard math symbols
\usepackage{nicefrac}       % compact symbols for 1/2, etc.
\usepackage{microtype}      % microtypography
\usepackage{xcolor}         % colors
\usepackage{amsmath}
\usepackage{graphicx}

\usepackage{makecell}   % For line breaks within cells
\definecolor{grayrow}{gray}{0.9}
\definecolor{lightgray}{gray}{0.92}

\usepackage[table]{xcolor}

\title{YingMusic-Singer: Zero-shot Singing Voice Synthesis and Editing with Annotation-free Melody Guidance}

\author{
\textbf{Junjie Zheng}$^{1}$ \quad
\textbf{Chunbo Hao}$^{1,2}$ \quad
Guobin Ma$^{1,2}$ \quad
Xiaoyu Zhang$^{1,3}$ \\
\textbf{Gongyu Chen}$^{1}$ \quad
\textbf{Chaofan Ding}$^{1}$ \quad
\textbf{Zihao Chen}$^{1}$ \quad
\textbf{Lei Xie}$^{2}$ \\[4pt]
$^{1}$AI Lab, GiantNetwork \quad
$^{2}$ASLP Lab, Northwestern Polytechnical University \\
$^{3}$University College London\quad
}

\begin{document}

\maketitle

\begin{abstract}
Singing Voice Synthesis (SVS) remains constrained in practical deployment due to its strong dependence on accurate phoneme-level alignment and manually annotated melody contours—requirements that are resource-intensive and hinder scalability. To overcome these limitations, we propose a melody-driven SVS framework capable of synthesizing arbitrary lyrics following any reference melody, without relying on phoneme-level alignment. Our method builds on a Diffusion Transformer (DiT) architecture, enhanced with a dedicated melody extraction module that derives melody representations directly from reference audio. To ensure robust melody encoding, we employ a teacher model to guide the optimization of the melody extractor, alongside an implicit alignment mechanism that enforces similarity distribution constraints for improved melodic stability and coherence. Additionally, we refine duration modeling using weakly-annotated song data and introduce a Flow-GRPO reinforcement learning strategy with a multi-objective reward function to jointly enhance pronunciation clarity and melodic fidelity. Experiments show that our model achieves superior performance over existing approaches in both objective measures and subjective listening tests, especially in zero-shot and lyric adaptation settings, while maintaining high audio quality without manual annotation. This work offers a practical and scalable solution for advancing data-efficient singing voice synthesis. To support reproducibility, we release our inference code and model checkpoints. Code and weights are available at: \url{https://github.com/GiantAILab/YingMusic-Singer}
\end{abstract}

\section{Introduction}

The digital entertainment industry has been continuously evolving, driven by advancements in audio and music technologies. Among these, Singing Voice Synthesis (SVS) stands out as a pivotal research direction with substantial application potential in music production, virtual singers, personal creative endeavors, and interactive media \cite{chen2020hifisinger, zhang2022wesinger, zhang2022visinger, hong2023unisinger}. Compared to general speech synthesis, SVS must simultaneously satisfy requirements for clear speech content, accurate melodic pitch, and natural singing expression, rendering it considerably more complex than speech synthesis model\cite{cho2021surveyrecentdeeplearningdriven}. Conventional SVS methods have long relied on precise 
phoneme-level duration and pitch annotations during both training and inference stages \cite{lu2020xiaocicesing}. This dependency not only necessitates specialized data production pipelines but also impedes the acquisition of large-scale training data, thereby significantly hindering the widespread adoption and industrial deployment of SVS technology. Recently, there has been a growing demand in the industry for melody-controlled singing voice synthesis, where a vocal is generated to match a given reference melody. This approach requires only the lyric and a reference melody audio as inputs, enabling users without professional musical expertise to participate in creative activities—rather than restricting music creation to trained professionals.

However, existing singing voice synthesis methods still exhibit considerable limitations, resulting in a notable performance gap between real-world applications and expectations. On one hand, the vast majority of systems depend on manual labor or alignment tools to obtain precise MIDI rhythms and phoneme-level duration annotations \cite{zhang2022m4singer,huang2021multi,wang2022opencpop,hong2024texttosongcontrollablemusicgeneration}. Such annotations are prohibitively expensive and difficult to scale to songs of arbitrary styles or languages. On the other hand, current approaches typically support only fixed lyric-melody pairs as seen during training. When users attempt to substitute lyrics, mix languages, or alter musical syntactic structures, the inherent mismatch between phoneme counts and melodic beats often leads to issues such as robotic pronunciation, rhythmic misalignment, and unnatural phrasing, substantially degrading the auditory experience. Furthermore, most contemporary methods lack zero-shot capability; their performance deteriorates significantly when encountering unseen text or prosodic structures, which contradicts practical application needs. Thus, transitioning from "usable" to "easy-to-use and practical" remains a critical bottleneck for SVS deployment.

To address these challenges, we propose a singing voice synthesis system that eliminates phoneme-level duration and pitch annotations, and enables free combination of arbitrary lyrics with any reference melody. We design a generative model based on the Diffusion Transformer (DiT) \cite{dhariwal2021diffusion}, incorporating a melody extraction module to directly derive melody information from reference songs, which is then used as a melodic condition during synthesis to avoid reliance on manual annotations. Recognizing that merely conditioning on melody may not ensure structural adherence to the reference track's overall melodic progression, we further introduce a implicit guidance mechanism based on similarity distribution constraints. Specifically, we compute similarity distribution matrices for both the reference song's MIDI and the model's acoustic flow representations derived via flow-matching, progressively minimizing their discrepancy during training to enable the model to more accurately follow the structural characteristics of the reference melody. This design yields significantly improved melodic stability and singing coherence compared to traditional conditional control methods.

To tackle the issue of duration mismatch between lyrics and melody, we optimize the model using training datas with just sentence-level timestamps \cite{diffrhythm}, allowing it to automatically infer reasonable duration allocations without phoneme alignment, thereby mitigating problems such as lyric squeezing, beat drift, and abrupt phrasing. Additionally, we pioneer the integration of reinforcement learning into the DiT-based SVS task. Through Flow-GRPO \cite{liu2025flowgrpo} policy fine-tuning, we construct a multi-objective reward function that incorporates content accuracy and melodic accuracy, leading to simultaneous improvements in both objective metrics and subjective listening quality.

Our principal contributions are summarized as follows:

\begin{itemize}
    \item \textbf{End-to-End Melody-Driven SVS System for Real-World Applications:} We propose a system that can synthesize arbitrary lyrics with any reference melody without requiring precise phoneme-level duration or pitch annotations. The model automatically learns to align lyrics with melody, substantially reducing both data production costs and usage barriers while improving real-world applicability.

    \item \textbf{Annotation-free Melodic Guidance Based on DiT and Weak Alignment Optimization:} To achieve effective alignment between lyrics and melody, we integrate a melody extraction module to obtain melodic features under teacher guidance, alongside a similarity distribution matrix constraint. This approach ensures that the generated acoustic flow adheres structurally to the reference melody, leading to enhanced pitch stability and improved vocal naturalness.
    
    \item \textbf{Reinforcement Learning Post-Training via Flow-GRPO:} We incorporate reward models that consider both content accuracy and melodic similarity, and further enhance synthesis quality through policy optimization. After post-training, the model demonstrates improvement across all relevant metrics compared to the base model, with particularly stronger performance in zero-shot and lyric editing scenarios. These results validate the effectiveness of the proposed post-training approach.
\end{itemize}

Experimental results demonstrate that our method significantly outperforms existing systems across multiple experimental setup, exhibiting more stable melodic control and more natural singing expressiveness, especially in lyric modification and zero-shot synthesis settings. Moreover, even without manual alignment annotations, our approach maintains auditory quality comparable to — or even better than — models trained with precise alignment, confirming the practicality of our solution for real-world applications.

\section{Related Work}

\subsection{Singing Voice Synthesis (SVS)}
Singing Voice Synthesis (SVS) aims to generate natural, fluent, and pitch-accurate singing voices based on lyrics and melody. Early systems (e.g., XiaoiceSing \cite{lu2020xiaocicesing}, VISinger \cite{zhang2021visinger}) relied on precise phoneme alignment and manually annotated MIDI information, achieving high-quality singing synthesis through a two-stage acoustic model-vocoder structure, but suffering from high training costs and difficult data production. With the introduction of diffusion models, works like DiffSinger \cite{liu2022diffsinger} and SmoothSinger \cite{sui2025smoothsinger} have made significant improvements in sound quality and stability, making end-to-end diffusion-based SVS a mainstream direction. In recent years, the research focus has gradually shifted towards zero-shot and cross-lingual capabilities. TCSinger 2 \cite{zhang2025tcsinger2} achieves zero-shot style transfer (not voice cloning) through a fuzzy boundary content encoder and Flow-Transformer structure, supporting multilingual and multi-style controllable singing; CoMeLSinger \cite{zhao2025comelsinger} models lyrics and pitch using discrete tokens and achieves structured melody control through contrastive learning; Transinger \cite{shen2025transinger}, based on an IPA phonetic decomposition strategy, shows good generalization in unseen language scenarios. Additionally, RMSSinger \cite{he2023rmssinger} and N-Singer \cite{lee2022nsingernonautoregressivekoreansinging} attempt to reduce alignment dependency and improve efficiency under real musical scores and non-autoregressive frameworks. Although these methods have made progress in sound quality, stability, and multilingual generalization, they still generally rely on manually annotated training data, lack zero-shot voice cloning capability, and require external conditional inputs such as MIDI or pitch sequences during inference. This makes it difficult for systems to be widely used in non-professional scenarios. Therefore, we propose an architecture that does not belong to the traditional SVS paradigm: it can generate natural singing voices without inputting precise phoneme-level duration or pitch annotations, supports zero-shot voice cloning, and is compatible with traditional SVS tasks, achieving a leap from "usable" to "easy-to-use and effective".

\subsection{Reinforcement Learning (RL)}
In recent years, reinforcement learning (RL) has been widely used for alignment and generation quality optimization in large models. Since RLHF was proposed for summarization and instruction-following tasks \cite{stiennon2020learning, ouyang2022training}, preference modeling based on PPO became the standard paradigm \cite{stiennon2020learning, ouyang2022training, schulman2017proximal}. Subsequently, Group Relative Policy Optimization (GRPO) significantly reduced the complexity of RL fine-tuning by using within-group relative scores instead of an explicit value function \cite{shao2024deepseekmath}, and has been applied to Flow Matching and Rectified Flow models, such as Flow-GRPO and Dance-GRPO, effectively improving compliance and aesthetic quality in text-to-image generation tasks \cite{liu2025flowgrpo, xue2025dancegrpo}. In the speech domain, existing work has introduced RLHF into emotional or diffusion-based speech synthesis tasks (e.g., i-ETTS \cite{liu2021reinforcement}, DLPO \cite{chen2024dlpo}). The recent F5R-TTS \cite{sun2025f5r} further applied GRPO policy optimization to a DiT backbone for TTS tasks, achieving end-to-end reward-driven speech generation \cite{zhang2025f5rtts}. However, there is still a lack of research systematically applying reinforcement learning to SVS. Existing methods generally rely on manual annotations and external melodic input, making it difficult to directly optimize singing quality through reward signals. To this end, we propose a multi-objective reward model integrating content accuracy and melodic similarity, and improve synthesis performance based on Flow-GRPO policy optimization. Across multiple experimental setup, our model outperforms existing systems, particularly excelling in zero-shot and lyric modification scenarios.

\section{Method}

\subsection{Overview}

\begin{figure*}
    \centering
    \includegraphics[width=\linewidth]{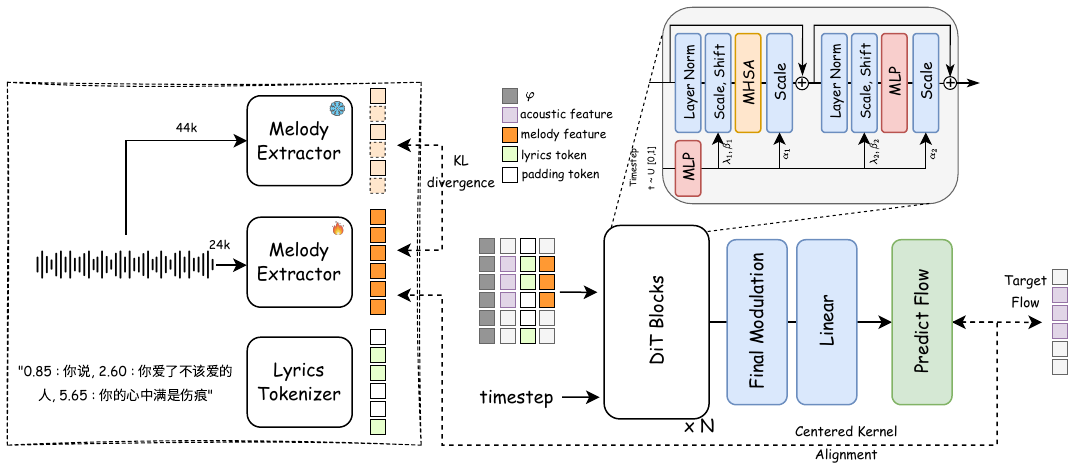}
    \caption{Architecture overview of SVS model}
    \label{fig:architecture}
\end{figure*}

This paper presents an end-to-end singing voice synthesis (SVS) framework that generates singing voices with high linguistic accuracy and melodic consistency relative to the raw input audio melody and textual lyrics. Unlike conventional pipeline models that treat melody extraction and synthesis as separate stages, our method adopts a synergistically integrated architecture. As shown in Figure \ref{fig:architecture}, the framework consists of two tightly coupled components: (1) an Online Melody Extraction Module, which extracts frame-level melodic representations directly from the input audio via a parameterized extractor; and (2) a Diffusion Transformer-based SVS Module, implemented as a conditional denoising diffusion model that synthesizes the final singing voice conditioned on the audio prompt, lyrics, and the extracted melodic representation.

To ensure effective melodic conditioning and joint module enhancement, the framework incorporates two key mechanisms. First, we introduce a distillation-based joint optimization strategy, where the melody extractor is trained end-to-end together with the synthesis module. A frozen, pre-trained teacher melody model supplies stable supervisory signals by minimizing divergence, allowing the online extractor to adaptively refine its representations for improved downstream synthesis—thereby promoting mutual performance gains. Second, to ensure that melodic conditions are meaningfully utilized during generation, we impose a representation-layer alignment constraint based on Centered Kernel Alignment (CKA) \cite{cka}. This constraint explicitly maximizes the correlation between the extracted melodic representations and the internal features of the synthesis model, strengthening melodic guidance and ensuring high consistency between the generated singing and the input melody.

Furthermore, we integrate reward models that assess content accuracy and melodic similarity, and employ policy optimization to further refine synthesis quality. After post-training, the model demonstrates improvements across all relevant metrics compared to the base model, with particularly notable gains in zero-shot and lyric editing scenarios. Through online joint optimization, internal representation alignment, and reward-guided fine-tuning, our approach ensures accurate and efficient flow of melodic information from input to output, leading to high-quality and robust singing voice synthesis.

\subsection{Pre-training}

\paragraph{Online Melody Learning and Joint Optimization.}
Traditional singing synthesis pipelines often treat melody extraction as an independent, fixed pre-processing step, which can lead to error propagation through the pipeline, and the extracted melodic features may deviate from the learning objective of the downstream synthesis model. To solve this problem, we design an online learning melody extractor and perform joint optimization with the singing synthesis model. Specifically, our melody extractor $E_{\phi}$ is an encoder network with parameters $\phi$. It takes preprocessed raw audio $x$ as input and outputs a frame-level melody representation sequence $m_e = E_{\phi}(x)$, where $m_e \in \mathbb{R}^{T \times D_m}$, $T$ is the number of time frames, and $D_m$ is the dimension of the melody representation.

To ensure that this learning extractor captures realistic and effective melody information, we introduce a distillation constraint based on KL divergence. We employ a teacher model\footnote{https://github.com/openvpi/SOME} $E_{teacher}$, pre-trained on a small-scale music dataset with accurate MIDI annotations and then frozen, to provide stable melody supervision \cite{SOME}. The melody representation extracted by this teacher model, $m_{teacher} = E_{teacher}(x)$, is treated as a "soft label". We guide the learning process of the student extractor $E_{\phi}$ by minimizing the Kullback-Leibler divergence \cite{kullback1951information} between the output of the student extractor $E_{\phi}$ and the teacher output $m  _{teacher}$. This constraint loss function is defined as follows:
\begin{equation}
\mathcal{L}_{KD} = D_{KL}(\text{Proj}(m_e) \| m_{teacher})
\end{equation}
where $\text{Proj}(\cdot)$ is a projection layer used to align the dimension of $m_e$ with that of $m_{teacher}$. This loss function ensures that the student model, while maintaining flexibility, produces melodic semantics consistent with a relatively accurate pre-trained model.

During joint training, the parameters $(\phi, \theta)$ of the melody extractor $E_{\phi}$ and the singing synthesis model $G_{\theta}$ are updated together. The training loss of the singing synthesis model (the denoising loss of the diffusion model) provides direct gradient feedback to $E_{\phi}$ regarding "what kind of melodic features are beneficial for the synthesis task". This design enables the melody extractor to adaptively optimize its representations, no longer merely pursuing general melody extraction accuracy but focusing on the melodic features that most contribute to singing generation, thereby enhancing both its own performance and the end-to-end performance of the entire system.

\paragraph{Melody-Content Alignment Constraint Based on CKA.}
In conditional generation models, ensuring high correlation between the generated content and the given condition is crucial. To further strengthen the guiding role of the melodic condition for the generated song, we introduce a CKA loss \cite{cka} to explicitly constrain the correlation between the internal representations during song synthesis and the input melody representation. We use a flow matching-based model as the backbone of the song synthesis model $G_\theta$, which learns a highly structured latent space when processing data. Let $z_l$ be the feature representation of an intermediate layer in the flow model, which encodes the semantic and acoustic information of the song being generated.

CKA is a reliable metric for measuring the similarity between two different representation spaces. We use linear CKA to measure the correlation between the melody representation $m_e$ and the flow model's internal feature $z_l$. Given two feature sets $m_e$ and $z_l$, their linear CKA is calculated as follows: (1) Compute covariance matrices: $K = m_e m_e^T$ and $L = z_l z_l^T$. (2) Center the covariance matrices. (3) The CKA value is given by the normalized form of the Hilbert-Schmidt Independence Criterion:
\begin{equation} 
\text{CKA}(K, L) = \frac{\|K^T L\|_F^2}{\|K^T K\|_F \|L^T L\|_F}
\end{equation}
where $\|\cdot\|_F$ denotes the Frobenius norm. Our goal is to maximize the CKA value between $m_e$ and $z_l$, i.e., to minimize the following CKA loss:
\begin{equation} 
\mathcal{L}_{CKA} = 1 - \text{CKA}(m_e, z_l)
\end{equation}
The introduction of this loss function encourages the song synthesis model, during the generation process, to maintain high structural consistency between its internal data flow (corresponding to the timbre, rhythm, etc., of the song) and the externally provided melody condition. This is equivalent to imposing a correlation inductive bias within the model, effectively preventing the problem of the generated result deviating from the input melody, thereby significantly improving the accuracy and robustness of melodic guidance.

\paragraph{Overall Training Objective.}

The total training loss of our model is the weighted sum of the aforementioned losses:
\begin{equation} 
\mathcal{L}_{Total} = \mathcal{L}_{Diffusion} + \lambda_{KD} \cdot \mathcal{L}_{KD} + \lambda_{CKA} \cdot \mathcal{L}_{CKA}
\end{equation}
where $\mathcal{L}_{Diffusion}$ is the standard denoising score matching loss (mean squared error loss) of the diffusion model, and $\lambda_{KD}$ and $\lambda_{CKA}$ are hyperparameters used to balance the importance of each task.

\begin{figure}
    \centering
    \includegraphics[width=\linewidth]{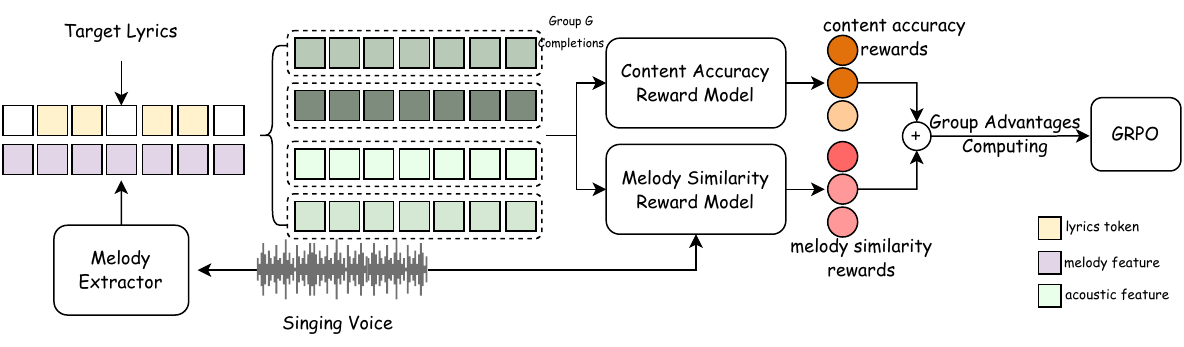}
    \caption{Multi-objective alignment for both intelligibility and melody similarity. This figure demonstrates how we utilize singing voice as melody prompts during post-training of YingMusic-Singer.}
    \label{fig:grpo}
\end{figure}

\subsection{Post Training}

\paragraph{Motivation.}

After pre-training, we observed that the model maintains high speaker similarity and melody similarity in zero-shot scenarios, though pronunciation clarity can occasionally be compromised. Moreover, in downstream applications, the model may encounter input distributions that differ from those seen during training. For instance, in singing voice synthesis (SVS), the model receives melody markers extracted from singing voices (as illustrated in Figure \ref{fig:grpo}) and integrates them with rhythmic cues derived from the input lyrics. In certain cases, the provided melody may not align well with the lyric content, requiring the model to adaptively balance adherence to the melody with faithfulness to the lyrics. Motivated by these considerations, this study introduces a post-training stage designed to enhance the model’s controllability over both lyrics and melody, as well as to improve its robustness to diverse input types. This stage employs the reinforcement learning-based GRPO method to further optimize the performance of the singing voice synthesis model (Figure \ref{fig:grpo}).

\paragraph{GRPO.}
This method focuses on strategy optimization. By designing and introducing reward functions for content accuracy and melody consistency, the model can iteratively update toward better singing performance during the post-training stage. This enables the model to preserve semantic clarity while maintaining consistent melodic structure, thereby producing higher-quality and more controllable singing outputs.

In this stage, we refine the model using reinforcement learning to directly optimize non-differentiable perceptual objectives. To enable stochastic policy optimization, we reinterpret the deterministic flow dynamics as a stochastic policy by injecting a small amount of noise into the ODE trajectory, similar to techniques used in Flow-GRPO \cite{liu2025flowgrpo}:
\begin{equation} 
d x_t = v_\theta(x_t, t)\, dt + \sigma_t\, d w_t,
\end{equation}
where $w_t$ is a standard Wiener process and $\sigma_t$ controls the injected stochasticity. We adopt a simple monotonic schedule $\sigma_t = a \sqrt{t / (1 - t)},$ where $a$ determines the overall noise level. To avoid the credit-assignment ambiguity associated with full SDE sampling, we inject randomness at a single uniformly sampled timestep while keeping all remaining steps deterministic \cite{team2025longcat}. 

For each prompt $c$, $G$ completions are generated and normalized rewards are used to compute the advantage $A^{(i)}$, leading to the training objective
\begin{equation}
\begin{aligned}
J(\theta)
&=\mathbb{E}_{c,\;t',\;\{x^i\}_{i=1}^G \sim \pi_{\mathrm{old}}}\!
\left[\frac{1}{G}\sum_{i=1}^{G}
\Big(r_{t'}^i(\theta)\, \hat{A}^i - 
\beta\, D_{\mathrm{KL}}\!\left(\pi_\theta\,\|\,\pi_{\mathrm{ref}}\right)_{t'}\Big)
\right],
\end{aligned}
\end{equation}
where $r_{t'}^i(\theta) $ is the policy ratio correcting for off-policy sampling, $t'$ denotes the uniformly sampled timestep at which stochasticity is injected, and $\pi_{\theta}$ and $\pi_{ref}$ represent the current policy and the frozen reference policy, respectively. This selective-noise scheme preserves exploration while substantially improving optimization stability.

\paragraph{Content Accuracy Reward.}

To assess the articulation clarity and content accuracy of the converted singing, we first employ an ASR model to compute a reward grounded in the word error rate ($\mathrm{WER}$). Specifically, given the transcription $\hat{y}$ of the generated singing and the corresponding reference text $y$, we calculate the $\mathrm{WER}$ as follows:
\begin{equation}
\mathrm{WER} = \frac{S+ D + I}{N},
\end{equation}
Specifically, $S$ denotes the number of substitution errors, $D$ denotes the number of deletion errors, and $I$ denotes the number of insertion errors, while $N$ is the total number of words in the reference text. To ensure that the reward correlates positively with better recognition outcomes, we define the content accuracy reward $R_{\mathrm{con}}$ as follows:

\begin{equation}
R_{\mathrm{con}} = 1 - \mathrm{WER}.
\end{equation}

\paragraph{Melodic Similarity Reward.}

We use the Pearson correlation coefficient between the generated pitch contour and the reference pitch contour (F0) as the melodic similarity reward. Specifically, we first extract the pitch trajectories of the two audio segments. Then, we compute the similarity only on voiced frames (i.e., frames with non-zero F0) to avoid interference from silence and invalid regions. For each sample, we denote the generated pitch sequence as $f^{(g)} = \{ f^{(g)}_1, f^{(g)}_2, \ldots, f^{(g)}_N \}$ and the target pitch sequence as $f^{(t)} = \{ f^{(t)}_1, f^{(t)}_2, \ldots, f^{(t)}_N \}$. The melodic similarity reward $R_{\mathrm{mel}}$ is then defined as the Pearson correlation between the two sequences:

\begin{equation}R_{\mathrm{mel}}(f^{(g)}, f^{(t)}) =\frac{\sum_{i=1}^N \left( f^{(g)}_i - \bar{f}^{(g)} \right) \left( f^{(t)}_i - \bar{f}^{(t)} \right)}{\sqrt{\sum_{i=1}^N \left( f^{(g)}_i - \bar{f}^{(g)} \right)^2\sum_{i=1}^N \left( f^{(t)}_i - \bar{f}^{(t)} \right)^2}}.
\end{equation}

The reward reflects the consistency of the melodic direction between the two audio segments. The higher the correlation coefficient, the closer the pitch change trend of the generated speech is to the target, thus obtaining a higher melodic similarity reward.

\paragraph{Multi-Objective Optimization.}

The final multi-objective reward for the sample $i$-th is:
\begin{equation}
R^i = \sum_k w_k R_k^i,
\end{equation}
and the group-relative advantage is:
\begin{equation}
\hat{A}^i = \frac{R^i - \mu}{\sigma},
\end{equation}
where $\mu$ and $\sigma$ denote the mean and standard deviation over the group for the same conditioning prompt, and $w_k$ denotes the weighting coefficient associated with the $k$-th reward term in the multi-objective formulation. This multi-objective RL framework enables direct optimization of lyric alignment and melody consistency, complementing the supervised objectives in preceding stages. In our implementation, the weights for both the melodic similarity reward and the content accuracy reward are set to $1$.

\begin{figure}
    \centering
    \includegraphics[width=0.6\linewidth]{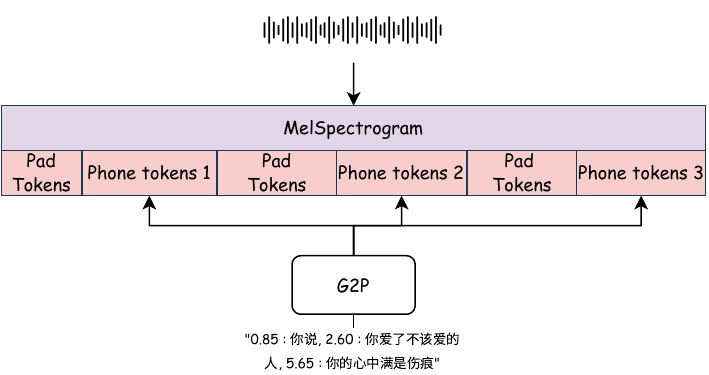}
    \caption{Lyrics go through G2P and are placed at the positions corresponding to their timestamps}
    \label{fig:lyicsPadding}
\end{figure}

\begin{table*}[ht]
\centering
\caption{Comparison of Singing Voice Synthesis and Editing tasks. Best results are highlighted in \textbf{bold}.}
\label{tab:results}
\resizebox{\textwidth}{!}{%
\begin{tabular}{clccccccc}
\toprule
\multirow{2}{*}{\textbf{Task}} & \multirow{2}{*}{\textbf{Method}} & \multicolumn{3}{c}{\textbf{Objective Metrics}} & \multicolumn{4}{c}{\textbf{Aesthetic Scores}} \\
\cmidrule(lr){3-5} \cmidrule(lr){6-9}
 & & WER (\%)$\downarrow$ & SIM (\%)$\uparrow$ & FPC (\%)$\uparrow$ & CE$\uparrow$ & CU$\uparrow$ & PC$\uparrow$ & PQ$\uparrow$ \\
\midrule
% Block 1: SVS (3 rows)
\multirow{3}{*}{\shortstack[c]{Zero-Shot\\Singing Voice\\Synthesis}} 
 & TCSinger \cite{tcsinger} & 3.47 & \textbf{94.41} & 77.79 & 5.56 & 6.20 & 1.77 & 7.36 \\
 & Vevo \cite{vevo} & 9.83 & 93.51 & \textbf{87.96} & 6.42 & \textbf{6.76} & \textbf{1.84} & \textbf{7.60} \\
 & Ours & \textbf{1.28} & 93.95 & 81.28 & \textbf{6.57} & 6.68 & 1.72 & 7.58 \\
\midrule
% Block 2: SVE (6 rows to cover sub-headers + data)
\multirow{6}{*}{\shortstack[c]{Singing Voice\\Editing}} 
 & \textbf{\textit{Lyrics Editing}} & & & & & & & \\
 & \quad Vevo \cite{vevo} & 29.89 & \textbf{95.87} & 83.47 & 6.28 & \textbf{6.66} & \textbf{1.78} & \textbf{7.54} \\
 & \quad Ours & \textbf{16.58} & 95.36 & \textbf{89.53} & \textbf{6.31} & 6.52 & 1.73 & 7.50 \\
 \cmidrule{2-9}
 & \textbf{\textit{Structural Editing}} & & & & & & & \\
 & \quad Vevo \cite{vevo} & 30.63 & \textbf{96.11} & 89.27 & 6.28 & \textbf{6.65} & \textbf{1.80} & \textbf{7.54} \\
 & \quad Ours & \textbf{18.44} & 95.47 & \textbf{90.34} & \textbf{6.38} & 6.47 & 1.75 & 7.50 \\
\midrule
% Block 3: Zero-shot SVE (6 rows)
\multirow{6}{*}{\shortstack[c]{Zero-Shot\\Singing Voice\\Editing}} 
 & \textbf{\textit{Lyrics Editing}} & & & & & & & \\
 & \quad Vevo \cite{vevo} & 67.31 & 93.46 & \textbf{83.91} & 6.32 & 6.64 & \textbf{1.83} & 7.54 \\
 & \quad Ours & \textbf{15.18} & \textbf{93.75} & 82.84 & \textbf{6.78} & \textbf{6.71} & 1.77 & \textbf{7.59} \\
 \cmidrule{2-9}
 & \textbf{\textit{Structural Editing}} & & & & & & & \\
 & \quad Vevo \cite{vevo} & 73.97 & 93.53 & \textbf{82.52} & 6.32 & \textbf{6.67} & \textbf{1.83} & 7.53 \\
 & \quad Ours & \textbf{12.62} & \textbf{93.77} & 81.19 & \textbf{6.54} & 6.66 & 1.75 & \textbf{7.57} \\
\bottomrule
\end{tabular}%
}
\label{table:com_baselines}
\end{table*}

\begin{table}[ht]
\centering
\small % Reduces font size (options: \footnotesize, \scriptsize)
\setlength{\tabcolsep}{3.5pt} % Reduces space between columns
\caption{Performance comparison on Zero-Shot Singing Voice Editing on subjective evaluation.}
\label{tab:zeroshot}
\begin{tabular}{lcc}
\toprule
\multirow{2}{*}{\textbf{Model}} & \multicolumn{2}{c}{\textbf{Zero-Shot Singing Voice Editing}} \\
\cmidrule(lr){2-3}
 & \textit{N-CMOS} & \textit{Melody-MOS} \\ % Replace with actual metric names
\midrule
Vevo & -0.75 $\pm$ 0.12 & 1.62 $\pm$ 0.38 \\
Ours & \textbf{0.00 $\pm$ 0.00} & \textbf{1.76 $\pm$ 0.35} \\
\bottomrule
\end{tabular}
\label{table:mos}
\end{table}

\begin{table}[ht]
\centering
\small % Reduces font size (options: \footnotesize, \scriptsize)
\setlength{\tabcolsep}{3.5pt} % Reduces space between columns
\caption{Ablation study results. Best results are highlighted in \textbf{bold}.}
\label{tab:ablation}
\begin{tabular}{lccccccc}
\toprule
\multirow{2}{*}{\textbf{Model}} & \multicolumn{3}{c}{\textbf{Objective Metrics}} & \multicolumn{4}{c}{\textbf{Aesthetic Scores}} \\
\cmidrule(lr){2-4} \cmidrule(lr){5-8}
 & WER (\%)$\downarrow$ & SIM (\%)$\uparrow$ & FPC (\%)$\uparrow$ & CE$\uparrow$ & CU$\uparrow$ & PC$\uparrow$ & PQ$\uparrow$ \\
\midrule
full & \textbf{15.18} & \textbf{93.75} & \textbf{82.84} & \textbf{6.78} & 6.71 & 1.77 & 7.59 \\
w/o post-training & 16.75 & 93.31 & 76.64 & 6.42 & 6.51 & \textbf{1.78} & 7.46 \\
w/o cka-alignment & 16.49 & 93.41 & 82.73 & 6.60 & \textbf{6.73} & 1.73 & \textbf{7.60} \\
\bottomrule
\end{tabular}
\end{table}

\section{Experiments}

\subsection{Implementation}

We initialize the parameters of our DiT-based decoder from a pre-trained DiT TTS model, F5-TTS \cite{f5tts}, to expedite convergence and improve generalization. During training, lyrics are padded following the DiffRhythm \cite{diffrhythm} strategy, as shown in Fig. \ref{fig:lyicsPadding}. At inference, rather than using fine-grained timestamps, we separate the prompt from the generated content with a single timestamp. Our DiT architecture adheres to that of F5-TTS, consisting of 12 decoder layers with a hidden size of 1024. It employs 16-head self-attention mechanisms with 64 dimensions per head, amounting to a total of 0.3B parameters. To enable classifier-free guidance (CFG), we apply 20\% dropout independently to both the lyrics, audio prompts and melodic condition. The diffusion process uses an Euler ODE solver with 32 sampling steps and a CFG scale of 2 during inference. In the post-training stage, the FireRedASR model~\cite{fireredasr} is utilized to compute content accuracy reward, whereas the RMVPE model \cite{Wei_2023} extracts pitch trajectories and calculates melodic similarity reward. All models are optimized with the AdamW optimizer using $\beta_1 = 0.9$ and $\beta_2 = 0.95$. A learning rate of $1 \times 10^{-4}$ is applied, with linear warm-up over the first 2k steps followed by linear decay for the rest of training. Training is conducted on 8 A800 80GB GPUs with a batch size of 116,000 audio frames (approximately 0.35 hours) for 110k steps. Furthermore, since the sampling rate of the input audio for the supervised pre-trained MIDI extractor (44.1 kHz) and the frame rate of the Mel spectrogram differ from those of our backbone network (24 kHz), we resample the features extracted by the online melody extractor when computing the KL divergence. The coefficient $\lambda_{KD}$ is fixed at $1$ throughout training, while $\lambda_{CKA}$ decays from $0.3$ to $0.01$ over the first 2.5k steps.

\subsection{Training Data}

For singing voice training, we employed the data preparation pipeline introduced in DiffRhythm \cite{diffrhythm} to curate a dataset consisting of 3.7K hours of Mandarin singing vocals. These vocals were extracted via source separation from publicly available songs collected from the Internet. All audio signals were converted to mono channels at a sampling rate of 24 kHz and segmented into clips of approximately 30 seconds each. (Note that, as the melody extraction teacher model requires a 44 kHz input, the audio supplied to this model was resampled accordingly.) To facilitate the proposed multi-objective alignment task, the dataset was further refined by filtering approximately 500 hours of high-quality audio based on a combination of metrics, including DNSMOS \cite{dnsmos}, word error rate (WER), and aesthetic score \cite{aesthetic}.

\subsection{Evaluation Data}

We construct the evaluation set under various settings. For singing voice synthesis, we randomly selected 60 audio clips from GTSinger \cite{gtsinger}, which covers a broad spectrum of singing techniques, styles, and timbre types. For the zero-shot setting, we randomly chose five speakers from in-the-wild singing data that were not included in the training set. To evaluate singing voice editing, we built two dedicated datasets: one for modifying lyrics while preserving the original lyrical structure and character count, and another for modifying lyrics with changes to both structure and character count. All lyric modifications were generated using DeepSeek-V3.2\footnote{https://chat.deepseek.com/}, with each original song containing at least three variations.

\subsection{Evaluation Metrics}

\paragraph{Objective Metrics}

For objective evaluation, we employ several metrics to assess key aspects of the generated singing voices: intelligibility via Word Error Rate (WER, \(\downarrow\)), speaker similarity (SIM, \(\uparrow\)), and F0 correlation (FPC, \(\uparrow\)) \cite{svcc-2023,amphion-svc,vevo,vevo2}. WER is computed using the FireRedASR\footnote{https://huggingface.co/FireRedTeam/FireRedASR-AED-L} model \cite{fireredasr}. For SIM, we measure the cosine similarity between WavLM-TDNN\footnote{https://huggingface.co/microsoft/wavlm-base-sv} speaker embeddings \cite{wavlm} extracted from the generated samples and the corresponding reference audio.

\paragraph{Subjective Metrics}

For subjective evaluation, we employ the Comparative Mean Opinion Score (CMOS, scaled from -2 to 2, \(\uparrow\)) to measure several perceptual attributes of the generated samples: naturalness (N-CMOS). We also utilize the Melody-MOS metric introduced in Vevo2 \cite{vevo2} (ranging from 1 to 3) to specifically assess melody-following capability, with the following scoring criteria: 1 indicates “unable to follow the melody,” 2 corresponds to “roughly following the melody contour,” and 3 represents “accurately following all melodic details.”

\subsection{Controllability in Zero-shot Singing Voice Synthesis and Editing}

To evaluate the controllability of YingMusic-Singer over content and melody, we conducted a series of tasks including zero-shot Singing Voice Synthesis (SVS) and singing voice editing (encompassing both structural and lyrical modifications). In the zero-shot SVS task—specifically defined here as zero-shot timbre transfer—the model is conditioned on target lyrics, melodic notation (e.g., MIDI \cite{tcsinger}), and a reference audio waveform. The objective is to generate a singing voice that adheres to the target content and notes while preserving the reference timbre. Uniquely, YingMusic-Singer renders MIDI-like information into a reference singing melody, effectively treating the task as a melody-to-singing synthesis.

Table \ref{table:com_baselines} summarizes the comparison between our model and two baseline systems, TCSinger \cite{tcsinger} and Vevo \cite{zhang2025vevo}. Across nearly all tasks, the post-trained YingMusic-Singer demonstrates superior performance in lyric and melody transcription, achieving the lowest Word Error Rate (WER) and competitive F0 Pearson Correlation (FPC). Subjective evaluations (Table \ref{table:mos}) further indicate that YingMusic-Singer achieves higher N-CMOS scores than Vevo, reflecting superior naturalness.  Although the FPC is slightly lower than that of Vevo, it still exceeds the 80 \% level, indicating a strong ability to follow the target melody. Furthermore, the model achieves a significantly lower WER, indicating that it successfully adheres to the melodic contour while maintaining superior content fidelity. This performance underscores an effective post-training strategy that balances the often-competing objectives of content accuracy, naturalness, and melody adherence.

Furthermore, in the structural editing task—where lyrics and overall sentence structures are significantly altered—both Vevo and YingMusic-Singer maintain low WER and strong F0 correlation. This suggests that in singing voice editing, preserving the melodic direction of phonetic units is more critical than strictly maintaining the original lyrical count or sentence structure.

\subsection{Effectiveness of Training Strategies}

To evaluate the contribution of different modules in YingMusic-Singer, we conducted ablation studies focusing on the cka-alignment mechanism and the post-training procedure. Specifically, beginning with the pre-trained YingMusic-Singer-base model, we introduced a post-trained version optimized using an intelligibility reward (measured by WER) and a Melody Similarity Reward (measured by FPC Score). For the zero-shot singing voice editing task, we designed two subjective evaluation metrics: given target lyrics, a target melody (extracted from another singing voice), and a generated singing voice, participants assessed whether the generated output accurately followed the lyrics (lyrics accuracy) and the melody (melody accuracy). Experimental results indicate that each module contributes positively to the final outcome, as summarized in Table \ref{tab:ablation}. Removing the post-training stage led to a decline in nearly all metrics (WER: 15.18\% → 16.75\%; FPC: 82.84\% → 76.64\%). Additional observations include that cka-alignment facilitates faster convergence to melody-related guidance during the early training phase. However, the corresponding loss weights must be gradually reduced during training; otherwise, although a strong F0 correlation is achieved, it results in increased WER. Notably, when both rewards are applied together, the model demonstrates not only improved melody-following capability but also enhanced text-following performance. We hypothesize that this improvement stems from reinforced melody modeling, which in turn promotes clearer pronunciation and higher intelligibility. These findings further validate the advantages of our proposed multi-objective post-training strategy. Overall, the model achieves a better balance between content accuracy and melodic adherence.
  
\section{Conclusion}

In this paper, we propose a novel framework for melody-driven singing voice synthesis designed to overcome scalability limitations by removing the need for precise phoneme-level alignment and manual annotation. By integrating a Diffusion Transformer with automatic melody extraction and Flow-GRPO reinforcement learning, the system attains enhanced melodic stability and pronunciation clarity using only weakly-aligned training data. Experimental evaluations demonstrate that our method surpasses existing baselines in both objective measures and subjective listening tests, notably in zero-shot lyric adaptation settings, thereby providing a practical pathway toward more accessible audio content creation. Based on these outcomes, future research will focus on three key directions. First, we will expand multilingual capability through unified phoneme representations to support high-quality cross-lingual synthesis. Second, we will pursue audio fidelity improvements by incorporating advanced neural vocoders and latent diffusion models, targeting high-sampling-rate, studio-level output quality. Finally, we aim to strengthen generalization and expressive control by scaling model training on diverse in-the-wild datasets and disentangling vocal attributes, enabling fine-grained manipulation of emotional and stylistic features across varied musical contexts.

\bibliographystyle{unsrtnat}
\bibliography{ref}

\end{document}